\documentclass[aps,prl,twocolumn,superscriptaddress,floatfix]{revtex4-2}
\usepackage{amsmath}
\usepackage{amsfonts}
\usepackage{amssymb}
\usepackage{graphicx}
\usepackage{cancel}
\usepackage[switch,columnwise]{lineno}
\usepackage{titlesec}
\usepackage{xcolor}
\usepackage{hyperref}
\usepackage[none]{hyphenat}

\makeatletter
\renewcommand{\section}{\@startsection{section}{1}{0mm}
  {-\baselineskip}{0.5\baselineskip}{\bf\leftline}}

\renewcommand{\subsection}{\@startsection{section}{1}{0mm}
  {-\baselineskip}{0.5\baselineskip}{\bf\leftline}}
\makeatother

\begin{document}

\title{Realizing exceptional points by Floquet dissipative couplings in thermal atoms}

\author{Zimo Zhang}%
\affiliation{State Key Laboratory of Quantum Optics and Quantum Optics Devices, Institute of Opto-electronics, Shanxi University, Taiyuan, Shanxi 030006, China}%

\author{Fengbo Zhang}%
\affiliation{State Key Laboratory of Quantum Optics and Quantum Optics Devices, Institute of Opto-electronics, Shanxi University, Taiyuan, Shanxi 030006, China}%

\author{Zhongxiao Xu}%
\email{xuzhongxiao@sxu.edu.cn}
\affiliation{State Key Laboratory of Quantum Optics and Quantum Optics Devices, Institute of Opto-electronics, Shanxi University, Taiyuan, Shanxi 030006, China}%
\affiliation{Collaborative Innovation Center of Extreme Optics, Shanxi University, Taiyuan, Shanxi 030006, China}%

\author{Ying Hu}%
\email{huying@sxu.edu.cn}
\affiliation{State Key Laboratory of Quantum Optics and Quantum Optics Devices, Institute of Laser Spectroscopy, Shanxi University, Taiyuan, Shanxi, China}%
\affiliation{Collaborative Innovation Center of Extreme Optics, Shanxi University, Taiyuan, Shanxi 030006, China}%

\author{Han Bao}%
\email{hanbao@uni-mainz.de}
\affiliation{QUANTUM, Johannes Gutenberg-Universit\"{a}t Mainz, 55128 Mainz, Germany}

\author{Heng Shen}%
\affiliation{State Key Laboratory of Quantum Optics and Quantum Optics Devices, Institute of Opto-electronics, Shanxi University, Taiyuan, Shanxi 030006, China}%
\affiliation{Collaborative Innovation Center of Extreme Optics, Shanxi University, Taiyuan, Shanxi 030006, China}%

\begin{abstract}
Exceptional degeneracies and generically complex spectra of non-Hermitian systems are at the heart of numerous phenomena absent in the Hermitian realm. Recently, it was suggested that Floquet dissipative coupling in the space-time domain may provide a novel mechanism to drive intriguing spectral topology with no static analogues, though its experimental investigation in quantum systems remains elusive. We demonstrate such Floquet dissipative coupling in an ensemble of thermal atoms interacting with two spatially separated optical beams, and observe an anomalous anti-parity-time symmetry phase transition at an exception point far from the phase-transition threshold of the static counterpart. Our protocol sets the stage for Floquet engineering of non-Hermitian topological spectra, and for engineering new quantum phases that cannot exist in static systems.
\end{abstract}

\maketitle

\begin{figure*}
\centering
\includegraphics[width=0.9\textwidth]{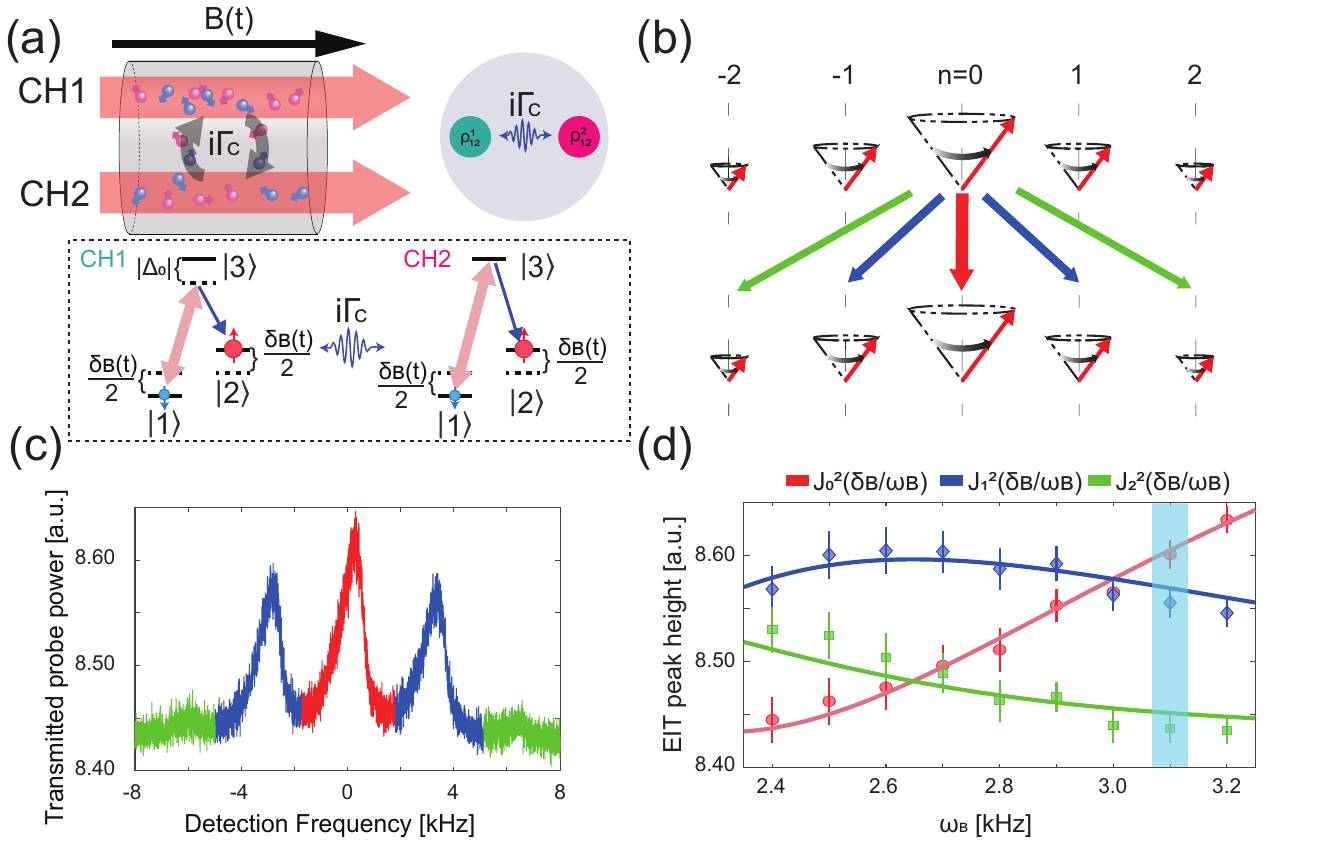}
\caption{\label{Fig:setup} Schematics for Floquet dissipative coupling in thermal atomic ensembles. (a) Experimental setup. Two spatially separated optical channels (CH1 and CH2) propagate in a paraffin-coated $^{87}\text{Rb}$ vapor cell at 54 $^{\circ} C$. The thermal motion of atoms mediates the dissipative coupling between two spin waves created in each channel via the $\Lambda$-type EIT interaction. The $|1\rangle$ and $|2\rangle$ are Zeeman sublevels of ground state $\left |5S_{1/2}, F=2 \right \rangle$, and $|3\rangle$ is the excited state$ \left |5P_{1/2}, F=1 \right \rangle$ of the $^{87}\text{Rb}$ D1 line. The left-circularly polarized control fields in Ch1 and Ch2 drive the dipole transition $|1\rangle\rightarrow|3\rangle$, with single photon detuning $\Delta_0$ and $0$, respectively, while the transition $|2\rangle\rightarrow|3\rangle$ is driven by right-circularly polarized probe. The cell resides inside a four-layer magnetic-shielding and is surrounded by a set of coils to generate a uniform bias magnetic field $B_0$ along the light propagation direction, inducing a Zeeman shift $\delta_0=17$ kHz. An oscillating magnetic field added to $B_0$ in \emph{z} direction leads to the periodic Zeeman splitting $\delta_B(t)=\delta_0+\delta_B\cos(\omega_Bt)$, thus creating Floquet sidebands of the spin wave in both channels. Different Floquet sidebands in two channels can be dissipatively coupled by tuning $\Delta_0$. (b) Illustration of Floquet dissipative coupling in two optical channels through sharing the spin waves. (c) A EIT spectra measured in CH2 under a periodic drive at $\omega_B=3.1$ kHz. It is marked with light blue area in (d). Since the control beam in CH2 is off, this signal carries the information of the spin wave transferred from CH1 to CH2. (d) EIT peak height as a function of the modulation frequency $\omega_B$. Solid lines show the numerical fit with Bessel functions $\mathcal{J}_m^2(\frac{\delta_B}{\omega_B})$ with ${\delta_B} \approx$ 3 kHz close to the estimated value (See supplementary materials for more details). In (b) and (c), both control and probe are on in CH1 while the control is off in CH2.}
\end{figure*}

Generically complex eigenvalues and the existence of exceptional points (EP) of non-Hermitian \cite{Bender} underlie exotic phenomena beyond the Hermitian framework \cite{Ueda,Sato,Coulais, Hu,Fan,Ma,Bergholtz,Ding}. Fascinating examples include the skin effect and anomalous bulk-edge relations~\cite{Lee,Kunst,Martinez Alvarez,Xiong,Yao,Xue}, exceptional nodal phases with open Fermi-Seifert surfaces \cite{Carlstrom2018,LeeLi,Carlstrom2019}, and topological energy transfer \cite{Xu}. 

The time-periodic modulation of a quantum system allows one to tailor interesting spectra and achieve new quantum phases of matter \cite{Eckardt,Dalibard,ZollerNP,Rudner,Lindner,HY,AndersonPRApp2016,MillerNJP2016,Jotzu,Aidelsburger}. In addition to providing powerful tools to engineer various spectra that could arise in static Hamiltonians, periodic driving also opens up the possibility of engineering new phases with no static analogues. In the non-Hermitian context, Floquet engineering has led to the recent realizations in photonic systems of a triple phase transition in a one-dimensional non-Hermitian synthetic quasicrystal \cite{Weidemann} and other intriguing phenomena. On the theoretical side, recent studies~\cite{Gong, An} conjectured that Floquet non-Hermitian systems featuring dissipative couplings in combined \textit{space-time} (Floquet) domain can host unique spectral topology and anomalous skin modes that cannot exist when the system is static. However, such Floquet dissipative coupling, an essential mechanism driving these novel phenomena, has yet to be investigated experimentally in quantum systems.

Here, we make an important step toward exploring the Floquet non-Hermitian engineering by reporting the first atomic realization of controllable dissipative couplings by periodic driving as a new approach for studying non-Hermitian phases and for dissipation engineering. Such a control has been previously challenging for thermal atoms, due to the difficulty to control the atomic diffusion induced dissipative couplings while still retaining sufficient atomic coherences. Using an oscillating magnetic field, we periodically shake a thermal atomic ensemble which interacts with two spatially separated laser beams in a vapor cell. This creates Floquet sidebands (each numbered by $n$) of atomic spinwaves in two spatially separated optical channels, which are dissipatively coupled via the ballistic atomic motion. We demonstrated our new approach via realizing tunable Floquet exceptional point (EP) in a wide parameter regime beyond the static setup, and, for the first time, in a setting where the synthetic (frequency) dimension is complemented by a real dimension. Such an ability to engineer non-local dissipations opens completely new possibilities of manipulating thermal atoms. For instance, it allows for the engineering of Floquet dissipative band structures and the realizations of ladders with a tunable number of legs, with important connections to non-Hermitian topological matter and dissipative Floquet phases that cannot exist in static systems. Compared with Floquet non-Hermitian systems in photonics and other platforms, which are classical in nature, atomic vapor system offers a unique platform for exploring the dissipation-mediated quantum correlations and fluctuations \cite{Cao,Sun}. And an example is illustrated in Ref. \cite{SM}.

We consider an anti-$\mathcal{PT}$ symmetric platform with thermal atoms, as depicted in Fig. \ref{Fig:setup} (a).  Our experiment consists of a paraffin-coated cylindrical cell, with diameter of 2.5 cm and length of 7.5 cm, containing isotopically enriched $^{87}$Rb vapor. Two spatially-separated optical beams (channels) emitted from a diode laser operating at the D1 line, labeled as CH1 and CH2, propagate along $z$-direction in an atomic vapor cell. In each channel, a strong right-circularly polarized control and left-circularly polarized weak probe, are completely overlapped and form the $\Lambda$-type electromagnetically induced transparency (EIT) configuration \cite{Harris,Fleischhauer,Novikova} immune to Doppler-broadening, creating a collective spin wave $\rho_{12}$ (ground-state coherence). A coherence-preserving environment, offered by the anti-relaxation coating inside the cell, ensures the dissipative coupling between spin waves in different channels at a rate $\Gamma_c$ via random atomic motion and wall bouncing. CH2's control is resonant with the transition $\left | 1 \right \rangle \rightarrow \left | 3  \right \rangle$ while a red detuning $\Delta_0<0$ is applied on CH1's control by using acoustic-optical modulators (AOMs). Both probe fields are nearly resonant with transition $\left | 2 \right \rangle \rightarrow \left | 3  \right \rangle$.

Given the $\Lambda$-type EIT in a configuration with Zeeman sublevels, we add an oscillating magnetic field $B_1\cos(\omega_Bt)$ to the static field $B_0$ along $z$-axis, leading to the periodically driven Zeeman levels. Floquet theory predicts a series of sidebands in frequency space with weight factors $\mathcal{J}_m(\frac{\delta_B}{\omega_B})$. Here $\mathcal{J}_m(x)$ is the Bessel function of the first kind of order $m$ with the modulation depth $\delta_B=\gamma B_1$ and the gyromagnetic ratio $\gamma$. As illustrated in Fig.~\ref{Fig:setup}(b), a time-periodic modulation forces a quantum state in each spatial channel dressed by all harmonics of the driving frequency, and the different harmonics- which manifests as the sidebands $(n=0, 1,2,...)$ of the atomic spin waves - corresponds to a synthetic dimension in the frequency (time) domain, in addition to the existing spatial dimension. As we show below, such a method could enable the creation and manipulation of a dissipative coupling between two arbitrary Floquet atomic sidebands emerged in two spatially separated optical channels in an unprecedented way inaccessible with static systems.

\begin{figure}
\centering
\includegraphics[width=0.4\textwidth]{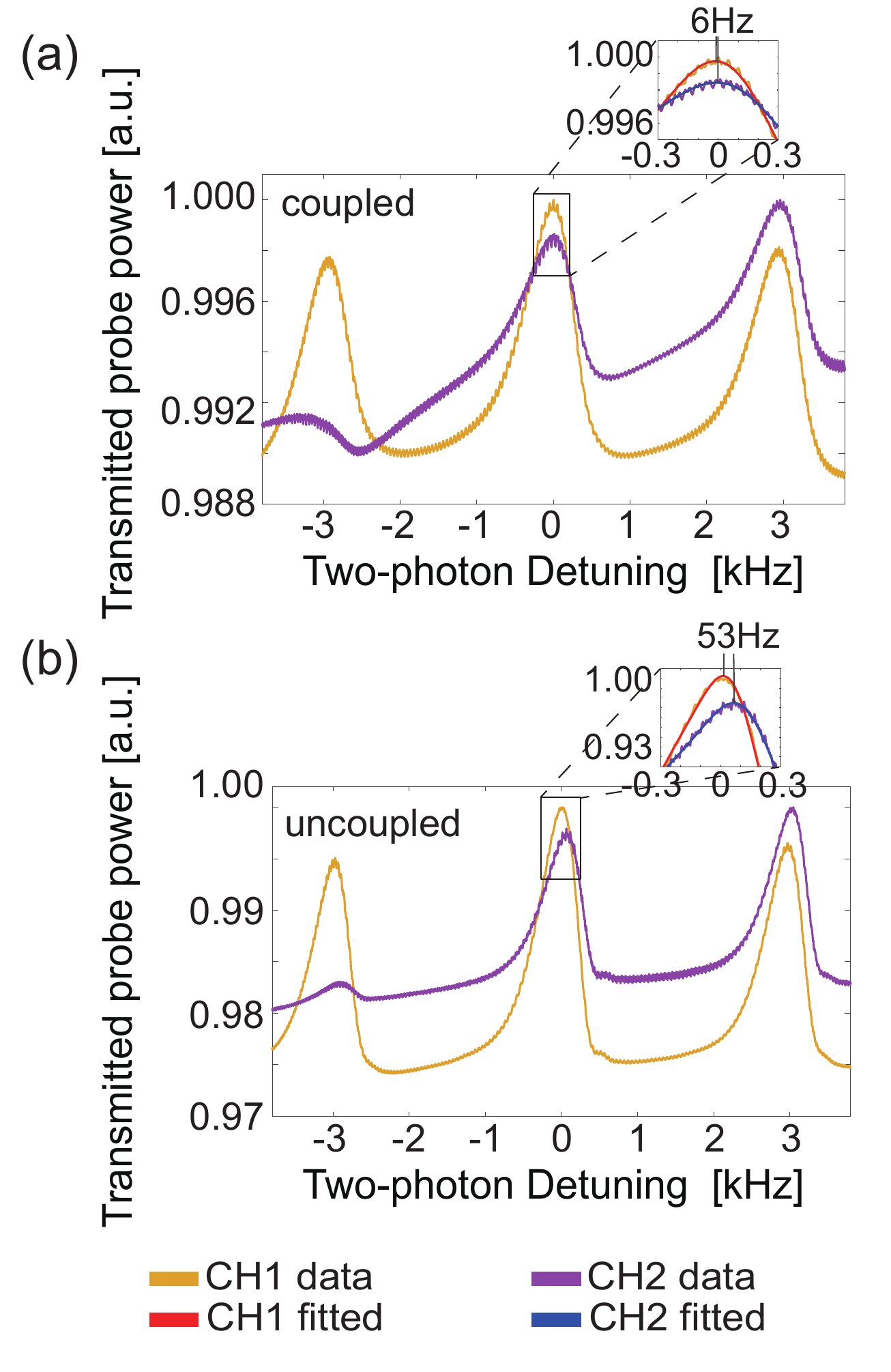}
\caption{\label{Fig2}Transmission spectra of output probe light in anti-$\mathcal{PT}$ symmetric optics under a time-periodic drive. Coupled and uncoupled EIT spectra are plotted in (a) and (b), respectively. Uncoupled EIT spectra with $\Gamma_c=0$  are separately measured from CH1 and CH2, by keeping the control and probe on in a single channel and fully blocking the light beams in the other one.  Here, red detuning $\left |\Delta_0  \right |=\text{3.05 \text{kHz}}\gg\Gamma_c$). The modulation frequency is set to be $\omega_B=3$ kHz, thus $|\Delta_0|-\omega_B=50$ Hz. Insets indicate the area of interest, corresponding to EIT carrier signal in CH1 and -\emph{1st}-order in CH2. Note that EIT spectra is shifted about 100 Hz due to the Stark effect. The experimental parameters here are control powers of 150 $\mu$W.}
\end{figure}

To gain some insights, recall that for the static setting with $B_1=0$, this system is governed by the non-Hermitian Hamiltonian $H=H_{\text{eff}}-i\gamma_{12} \mathbb{I}$ with~\cite{Xiao}
\begin{equation}\label{Hstatic}
H_{\text{eff}}=\begin{pmatrix}
\Delta_0& i\Gamma_c\\
i\Gamma_c&0
\end{pmatrix}
\end{equation}
and $\mathbb{I}$ is a $2\times 2$ identity matrix. Here, $\gamma_{12}$ is the common decay rate of the spin waves. $\Delta_0$ is the difference of single photon detuning for two channels as shown in Fig. \ref{Fig:setup}(a). The eigenvalues of the anti--$\mathcal{PT}$ Hamiltonian (\ref{Hstatic}) are
\begin{equation}
\nu_{\pm}=\frac{\Delta_0}{2}\pm\sqrt{\frac{\Delta_0^2}{4}-\Gamma_c^2},
\end{equation}
which correspond to the two eigen-EIT supermodes. The Hamiltonian $H_{\text{eff}}$ exhibits anti-$\mathcal{PT}$ symmetry~\cite{Xiao,Cao}: Two eigen-EIT resonance centres coincide (but with different linewidths) in the anti-$\mathcal{PT}$ symmetry-unbroken regime where $\left |\Delta_0  \right |<2\Gamma_c$, while the resonances bifurcate when $\left |\Delta_0  \right |>2\Gamma_c$, a manifestation of theanti-$\mathcal{PT}$ symmetry breaking; a symmetry-breaking transition occurs at the EP \cite{Ge,Li,Jing} in the parameter space with  $\left |\Delta_0  \right |=2\Gamma_c$, where the two supermodes coalesce. Clearly, by tuning $\left |\Delta_0  \right |$ to be far away from the EP, the dissipative coupling fails to pull the two spin-wave frequencies together as a result of a reduced efficiency of mutual coherence stimulation.

However, by applying a periodic driving to the spin-waves in the two optical channels, an anti-$\mathcal{PT}$ phase transition can be induced even though the static counterpart is deep in the anti-$\mathcal{PT}$ symmetry-broken phase, i.e. $\left |\Delta_0  \right |\gg 2\Gamma_c$. As will be shown, by choosing the appropriate driving frequency and strength, we can realize the dissipative coupling between two spin waves associated with the Floquet band index $n_1$ and $n_2$ in CH1 and CH2, respectively, governed by an effective Hamiltonian
\begin{equation}\label{HF}
H_F=\begin{pmatrix}
\Delta_0& i\Gamma_\text{{eff}}e^{-i(\Delta_0-n\omega_B)t}\\
 i\Gamma_\text{{eff}}e^{i(\Delta_0-n\omega_B)t}&n\omega_B
\end{pmatrix}
\end{equation}
with $n=n_1-n_2$. The Floquet dissipative coupling rate $\Gamma_\text{{eff}}>0$ is a function of $\Gamma_c$, $\delta_B$ and $\omega_B$. Equation (\ref{HF}) predicts an EP at $|-\Delta_0-n\omega_B|=2\Gamma_{\text{eff}}$. Hamiltonian (\ref{HF}) exhibits the eigenvalues
\begin{equation}\label{EF}
\nu^F_{\pm}=\frac{\Delta_0+n\omega_B}{2}\pm\sqrt{\frac{(\Delta_0-n\omega_B)^2}{4}-\Gamma_\text{eff}^2}.
\end{equation}

As a result, even in the regime $\left |\Delta_0  \right |\gg2\Gamma_c$, where the static system is far from the phase boundary, Floquet engineering provides a tool to realize the EP and induce the phase transition, by choosing the appropriate value of $n\omega_B$.

\begin{figure}
\centering
\includegraphics[width=0.45\textwidth]{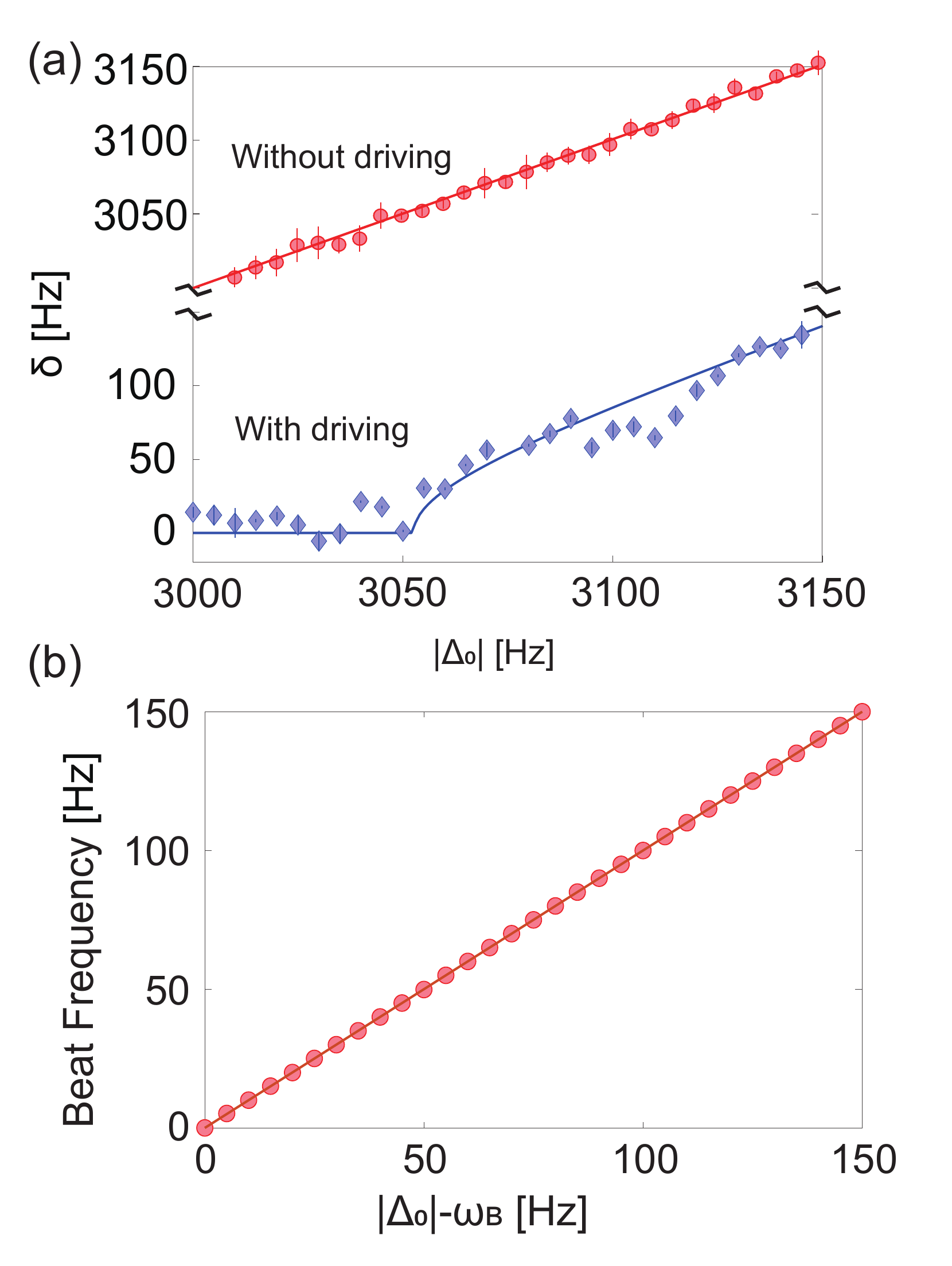}
\caption{\label{Fig3}Anomalous anti-$\mathcal{PT}$ symmetry phase transition assisted by Floquet dissipative coupling. (a) Measured EIT separation $\delta$ at varying $\left |\Delta_0  \right |$ with Floquet band index $n_1=1$ and $n_2=0$, far from the static  phase-transition threshold $\Delta_0$ = 2$\Gamma_c$. As a comparison, the blue curve represents the EIT peak separation in the static counterpart with $B_1=0$. (b) The beat frequency as a function of $|\Delta_0|-\omega_B$. The input power of the control beam in both channels is 250 $\mu$W in the experiment.}
\end{figure}

To demonstrate the realization of Floquet sidebands, we firstly create the ground state coherence $\rho_{12}$ in CH1 only, while probing the dissipatively-coupled spin wave due to the atomic motion in CH2 with the control beam off. Fig.\ref{Fig:setup}(c) showcases one example of the EIT spectra at $\omega_B=3.1$ kHz. Red, blue and green peaks correspond to carrier, first and second order of Floquet bands dissipatively transported from CH1, respectively. As shown in Fig. \ref{Fig:setup}(d), at various modulation frequencies  $\omega_B$, we extract the multiple EIT peak height, which are well fitted with a function $\propto\alpha\mathcal{J}_m^2(\frac{k}{\omega_B})$ ($m=0, 1, 2$) correspondingly. We thereby conjecture that the Floquet coupling rate is proportional to $\left |\mathcal{J}_m(\frac{\delta_B}{\omega_B})\right | $ if the $m$-th order band participates ~\cite{SM}.

We proceed to identify the presence of the Floquet dissipative coupling between collective spin waves associated with sideband index $n_1$ and $n_2$, respectively in CH1 and CH2, which is guaranteed by switching on both control and probe in each optical channel.  We consider the simplest case with $n_1-n_2=1$. We consider the case of $\Delta_0=-3.05$ kHz: although the associated static system is in the symmetry breaking phase, by adding a periodic drive with $\omega_B\approx \left |\Delta_0  \right |$, we observe the Floquet dissipative coupling efficiently between EIT carrier signal in CH1 (yellow) and -\emph{1st}-order signal in CH2 (violet), as displayed in Fig.\ref{Fig2}(a). As a benchmark, with only one channel on, the EIT spectra of uncoupling situation is measured and shown in Fig.\ref{Fig2} (b). Their centers are offset from each other, whose separation $\delta$ is almost the same as the detuning $\left |\Delta_0  \right |-\omega_B=50$ Hz. As shown in Fig.\ref{Fig2}(a), with both channels on, due to the Floquet dissipative coupling two EIT spectra with $n_1-n_2=1$ overlap, a manifestation of the synchronization of  spin in different sidebands due to atomic motion.

\begin{figure*}
\centering
\includegraphics[width=0.9\textwidth]{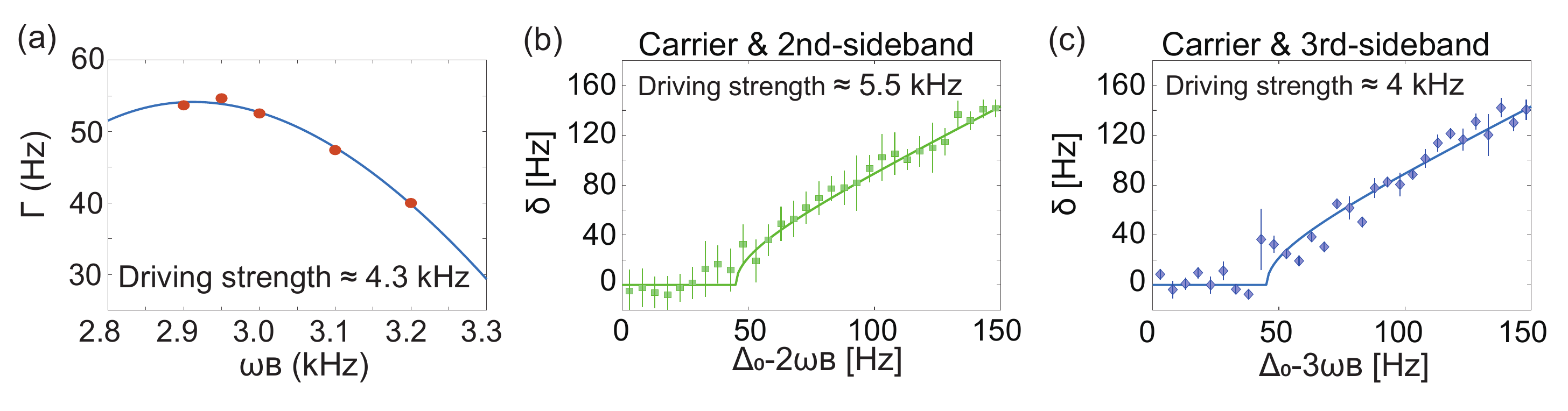}
\caption{\label{Fig4}Realizing the highly controlled Floquet dissipative coupling. (a) Dissipative coupling rate $\Gamma_{\text{eff}}$ as a function of modulation frequency $\omega_B$.  Solid line shows the numerical fit with a function $\propto\left |\mathcal{J}_0(\frac{\delta_B}{\omega_B})\mathcal{J}_1(\frac{\delta_B}{\omega_B})\right |$, ${\delta_B} \approx$ 4.3 kHz. The data points are obtained from the EP location. (b) Floquet assisted anti-$\mathcal{PT}$ symmetry phase transitions mediated by the dissipative coupling between different Floquet sidebands [$n_1=2; n_2=0$ in (b)] and [$n_1=3; n_2=0$ in (c)]. Proper driving strength are chosen to ensure the similar EP locations [$\Gamma_{\text{eff}}$=43 Hz in (b) and $\Gamma_{\text{eff}}$=45 Hz in (c)] for the dissipative coupling between two Floquet sidebands $n_1$ and $n_2$ with $n_1-n_2=n$.}
\end{figure*}

Moreover, by varying $\left |\Delta_0  \right |$, the evolution of Eq. \ref{HF}'s supermodes in the parameter space is revealed. Fig. \ref{Fig3}(a) shows the EIT peak separation with $n_1=1$ and $n_2=0$ as a function of $\left |\Delta_0  \right |$. Remarkably, we observe an anomalous transition between the symmetry unbroken phase and the symmetry breaking phase at $\left |\Delta_0  \right |=\omega_B+50$ Hz with $\omega_B=3$ kHz, although the static counterpart is deep in the broken phase (red curve).

In addition, we observe the the expected oscillating term $e^{-i(\Delta_0-n\omega_B)}$ in Eq. \ref{HF} which is examined by reading out frequencies of beating pattern in the EIT spectra with various $\Delta_0$ but a fixed $\omega_B$. This beating results from the redistribution and mixing of the two spin waves precessing, respectively, with $e^{-i\Delta_0}$ and $e^{-in\omega_B}$ in a common rotating frame coherently within the entire vapour cell. As illustrated in Fig. \ref{Fig3}(b), we find the beating frequency exactly equals to $\Delta_0-\omega_B$ with $n=1$.

Thanks to the high resolution of the phase-transition threshold \cite{Xiao}, $\Gamma_{\text{eff}}$ can be extracted from the EP location ($|-\Delta_0-n\omega_B|=2\Gamma_{\text{eff}}$) at the Hz level. Under a fixed modulation depth $\delta_B$, EPs locations are recorded at various modulation frequencies $\omega_B$. Fig. \ref{Fig4}(a) plots the deduced effective coupling rate $\Gamma_{\text{eff}}$ as a function of $\omega_B$, in a good agreement with the expected function (blue solid line)
\begin{equation}
\Gamma_{\text{eff}}=\left | \mathcal{J}_0(\frac{\delta_B}{\omega_B})\mathcal{J}_1(\frac{\delta_B}{\omega_B}) \right | \Gamma_c.
\end{equation}
This manifests that the ground-state coherence amplitude in each channel is tailored by
mutiplying $\left | \mathcal{J}_{n_i}(\frac{\delta_B}{\omega_B}) \right | $, and flying atoms carrying such spin coherence establish the dissipative coupling between two channels via thermal motion, corresponding to the bare rate $\Gamma_c$. $\Gamma_c$=93 Hz is extracted from the numerical  fit. More details can be found in  the supplementary material\cite{SM}. As the main result of this work, it allows us to manipulate the dissipative coupling with an addition degree of freedom, Floquet bands.

Having acquired the full information of Floquet dissipative coupling, we take a further step by applying to the $n$th-order Floquet EP transition in order to verify the capability of our approach. Essentially, given the initial detuning $\left |\Delta_0  \right |$ the $n$-th order process can be triggered by satisfying the condition $\omega_B=\left |\Delta_0  \right |/n$, wherein the +\emph{n}th-order collective spin wave in CH1 is dissipatively coupled to the carrier in CH2 through the thermal motion of atoms. Fig. \ref{Fig4}(b) and (c) shows the experimental observation of EP transitions for analog  two-, and three-photon process by setting the modulation frequency $\omega_B=$1.5 kHz and 1 kHz, respectively.  Remarkably, by choosing the proper modulation depth, i.e. modulation strength, we can ensure the same EP locations for the coupling of two Floquet sidebands with $n_1-n_2=n$, about $2\Gamma_{\text{eff}}=\left |\Delta_0  \right |-n\omega_B\approx  45$ Hz with $n=2, 3$, indicating the high controllability of Floquet dissipative coupling in a direct observation. Note that we directly observe Ch1 and Ch2's EIT peak locations experimentally, rather than the eigenmodes. Compared to the separation of the eigenmodes, directly observed frequency separation as a function of detuning $|\Delta_0|$ is less sharp at EP. More details can be found in Ref. \cite{SM}.

In summary, we have realized highly controllable Floquet dissipative couplings by periodically driving spatially separated thermal atomic ensembles. We show that a phase transition across an EP can be induced in an anti-$\mathcal{PT}$ system, initially deep within the symmetry broken phase. The locations of this Floquet EP, moreover, can be flexibly tuned. In contrast to recent works on exceptional points in a Floquet driven atoms or ions with tunable losses, which feature local dissipation e.g., local population loss \cite{Ion}, we realized the Floquet engineering of non-local dissipations in form of dissipative couplings for atoms, and demonstrated for the first time the possibility of easy engineering complex dissipative connectivity in a setting where the synthetic (frequency) dimension is complemented by a real dimension. At present, there are significant interests in using dissipative couplings as the building block for realizing various non-Hermitian phenomena \cite{Bardyn,Leefmans,Yoshida,Parto}. Our work not only offers an alternative route to the common approach based on static setups, it also paves the way for realizing non-Hermitian Floquet topological spectra and phases with no static analogues, including the Floquet EP geometry and skin effects as well as Floquet non-Hermitian topological arrays in the combined space-time domain, in quantum systems.

This work is supported by National Key R\&D Program of China under Grant No. 2020YFA0309400, NNSFC under Grant No. 12222409 and 12174081, 11974228, and the Key Research and Development Program of Shanxi Province (Grants No. 202101150101025). H. S. acknowledges financial support from the Royal Society Newton International Fellowship Alumni follow-on funding (AL201024) of the United Kingdom.


\begin{thebibliography}{9}\label{sec:TeXbooks}%
\bibitem{Bender}
C. M. Bender and S. Boettcher, Real spectra in non-Hermitian Hamiltonians having PT symmetry. \emph{Phys. Rev. Lett.} \textbf{80}, 5243-5246 (1998).
%
\bibitem{Ueda}
Z. Gong, Y. Ashida, K. Kawabata, K. Takasan, S. Higashikawa, and M. Ueda, Topological Phases of Non-Hermitian Systems. \emph{Phys. Rev. X} \textbf{8}, 031079 (2018).
%
\bibitem{Sato}
K. Kawabata, K. Shiozaki, M. Ueda, and M. Sato, Symmetry and Topology in Non-Hermitian Physics. \emph{Phys. Rev. X} \textbf{9}, 041015 (2019).
%
\bibitem{Coulais}
C. Coulais, R. Fleury, and J. van Wezel, Topology and broken Hermiticity. \emph{Nat. Phys.} \textbf{17}, 9-13 (2021).
%
\bibitem{Hu}
B. Hu, \emph{et al.} Non-Hermitian topological whispering gallery. \emph{Nature} \textbf{597}, 655-659 (2021).
%
\bibitem{Fan}
K. Wang, A. Dutt, C. C. Wojcik, \emph{et al.} Topological complex-energy braiding of non-Hermitian bands. \emph{Nature} \textbf{598}, 59-64 (2021).
%
\bibitem{Ma}
W. Wang, X. Wang, and G. Ma. Non-Hermitian morphing of topological modes. \emph{Nature} \textbf{608}, 50-55 (2022).
%
\bibitem{Bergholtz}
E. J. Bergholtz, J. C. Budich, F. K. Kunst, Exceptional topology of non-Hermitian systems.  \emph{Rev. Mod. Phys} \textbf{93}, 015005 (2021).
%
\bibitem{Ding}
K. Ding, C. Fang, and G. Ma, Non-Hermitian topology and exceptional-point geometries. \emph{Nat. Rev. Phys.} \textbf{4}, 745-760 (2022).
%
\bibitem{Lee}
T. E. Lee, Anomalous Edge State in a Non-Hermitian Lattice. \emph{Phys. Rev. Lett.} \textbf{116}, 133903 (2016).
%
\bibitem{Kunst}
F. K. Kunst, E. Edvardsson, J. C. Budich, and E. J. Bergholtz, Biorthogonal Bulk-Boundary Correspondence in Non-Hermitian Systems. \emph{Phys. Rev. Lett.} \textbf{121}, 026808 (2018).
%
\bibitem{Martinez Alvarez}
V. M. Martinez Alvarez, J. E. Barrios Vargas, and L. E. F. Foa Torres, Non-Hermitian robust edge states in one dimension:
Anomalous localization and eigenspace condensation at exceptional points. \emph{Phys. Rev. B} \textbf{97}, 121401 (2018).
%
\bibitem{Xiong}
Y. Xiong, Why does bulk boundary correspondence fail in some non-Hermitian topological models. \emph{J. Phys. Commun.} \textbf{2}, 035043 (2018).
%
\bibitem{Yao}
S. Yao, and Z. Wang, Edge States and Topological Invariants of Non-Hermitian Systems. \emph{Phys. Rev. Lett.} \textbf{121}, 086803 (2018).
%
\bibitem{Xue}
L. Xiao, T. Deng, K. Wang, G. Zhu, Z. Wang, W. Yi, and P. Xue. Non-Hermitian bulk–boundary correspondence in quantum dynamics. \emph{Nat. Phys.} \textbf{16}, 761-766 (2020).
%
\bibitem{Carlstrom2018}
J. Carlstr\"{o}m and E. J. Bergholtz, Exceptional links and twisted Fermi ribbons in non-Hermitian systems. \emph{Phys. Rev. A} \textbf{98}, 042114 (2018).
%
\bibitem{LeeLi}
C. H. Lee, G. Li, Y. Liu, T. Tai, R. Thomale, and X. Zhang, Tidal surface states as fingerprints of non-Hermitian nodal knot metals. arXiv:1812.02011.
%
\bibitem{Carlstrom2019}
J. Carlstr\"{o}m, M. St\aa lhammar, J. C. Budich, and E. J. Bergholtz, Knotted non-Hermitian metals,”  \emph{Phys. Rev. B} \textbf{99}, 161115 (2019).
%
\bibitem{Xu}
H. Xu, D. Mason, L. Jiang, and J. G. E. Harris, Topological energy transfer in an optomechanical system with exceptional points. \emph{Nature} \textbf{537}, 80-83 (2016).
%
\bibitem{Eckardt}
Andr\.{e} Eckardt, Atomic quantum gases in periodically driven optical lattices. \emph{Rev. Mod. Phys.} \textbf{89}, 011004 (2017).
%
\bibitem{Dalibard}
N. Goldman and J. Dalibard, Periodically Driven Quantum Systems: Effective Hamiltonians and Engineered Gauge Fields. \emph{Phys. Rev. X} \textbf{4}, 031027 (2014).
%
\bibitem{ZollerNP}
N. Goldman, J. C. Budich, and P. Zoller, Topological quantum matter with ultracold gases in optical lattices. \emph{Nat. Phys.} \textbf{12}, 639-645 (2016).
%
\bibitem{Rudner}
M. S. Rudner, N. H. Lindner, E. Berg, and M. Levin, Anomalous Edge States and the Bulk-Edge Correspondence for Periodically Driven Two-Dimensional Systems. \emph{Phys. Rev. X} \textbf{3}, 031005 (2013).
%
\bibitem{Lindner}
N. H. Lindner, G. Refae, abd V. Galitski, Floquet topological insulator in semiconductor quantum wells. \emph{Nat. Phys.} \textbf{7}, 490-495 (2011).
%
\bibitem{HY}
J. Carl Budich, Y. Hu, and P. Zoller, Helical Floquet Channels in 1D Lattices. \emph{Phys. Rev. Lett.} \textbf{118}, 105302 (2017).
%
\bibitem{AndersonPRApp2016}
D. A. Anderson, S. A. Miller, G. Raithel, J. Gordon, M. Butler, and C. Holloway, Optical Measurements of Strong Microwave Fields with Rydberg Atoms in a Vapor Cell, \emph{Phys. Rev. Appl.} \textbf{5}, 034003 (2016).
%
\bibitem{MillerNJP2016}
S. A. Miller, D. A. Anderson, and G. Raithel, Radio frequency-modulated Rydberg states in a vapor cell, \emph{New J. Phys.} \textbf{18}, 053017 (2016).
%
\bibitem{Jotzu}
G. Jotzu, M. Messer, R. Desbuquois, M. Lebrat, T. Uehlinger, D. Greif, and T. Esslinger, Experimental realization of the topological Haldane model with ultracold fermions. \emph{Nature} \textbf{515}, 237-240 (2014).
%
\bibitem{Aidelsburger}
M. Aidelsburger, M. Lohse, C. Schweizer, M. Atala, J. T. Barreiro, S. Nascimb\.{e}ne, N. R. Cooper, I. Bloch and N. Goldman, Measuring the Chern number of Hofstadter bands with ultracold bosonic atoms. \emph{Nat. Phys.} \textbf{11}, 162-166 (2015).
%
\bibitem{Weidemann}
S. Weidemann, M. Kremer, S. Longhi, and A. Szameitand, Topological triple phase transition in non-Hermitian Floquet quasicrystals. \emph{Nature} \textbf{601}, 354-359 (2022).
%
\bibitem{Gong}
X. Zhang and J. Gong, Non-Hermitian Floquet topological phases: Exceptional points, coalescent edge
modes, and the skin effect. \emph{Phys. Rev. B} \textbf{101}, 045415 (2020).
%
\bibitem{An}
H. Wu and J.-H. An, Floquet topological phases of non-Hermitian systems. \emph{Phys. Rev. B} \textbf{102}, 041119(R) (2020).
%
\bibitem{Cao}
W. Cao, X. Lu, X. Meng, J. Sun, H. Shen, and Y. Xiao, Reservoir-Mediated Quantum Correlations in Non-Hermitian Optical System. \emph{Phys. Rev. Lett.} \textbf{124}, 030401 (2020).
%
\bibitem{Sun}
J. Sun, X. Zhang, W. Qu, E. E. Mikhailov, I. Novikova, H. Shen, and Y. Xiao, Spatial Multiplexing of Squeezed Light by Coherence Diffusion. \emph{Phys. Rev. Lett.} \textbf{123}, 030401 (2019).
%
\bibitem{Harris}
K.-J. Boller, A. Imamo\v{g}lu, and S. E. Harris, Observation of electromagnetically induced transparency. \emph{Phys. Rev. Lett.} \textbf{66}, 2593 (1991).
%
\bibitem{Fleischhauer}
M. Fleischhauer, A. Imamoglu, and J. P. Marangos, Electromagnetically induced transparency: Optics in coherent media. \emph{Rev. Mod. Phys.} \textbf{77}, 633 (2005).
%
\bibitem{Novikova}
I. Novikova, R. L. Walsworth, and Y. Xiao, Electromagnetically induced transparency-based slow and stored light in warm atoms. \emph{Laser \& Photonics Review} \textbf{6}, 333-353 (2012).
%
\bibitem{Ge}
L. Ge and H. E. T{\"u}reci, Antisymmetric PT-photonic structures with balanced positive- and negative-index materials. \emph{Phys. Rev. A} \textbf{88}, 053810 (2013).
%
\bibitem{Li}
Y. Li \emph{et al.}, Anti-parity-time symmetry in diffusive systems. \emph{Science} \textbf{364}, 170 (2019).
%
\bibitem{Jing}
H. Zhang, R. Huang, S.-D. Zhang, Y. Li, C.-W. Qiu, F. Nori, and H. Jing, Breaking Anti-PT Symmetry by Spinning a Resonator. \emph{Nano Lett.} \textbf{20} 7594 (2020)
%
\bibitem{Xiao}
Peng, P. \emph{et al.} Anti-parity-time symmetry with flying atoms. \emph{Nat. Phys.} \textbf{12}, 1139-1145 (2016).
%
\bibitem{SM}
See Supplemental Material at http://link.aps.org/supplemental/XXX.
%
\bibitem{Xiao2}
D. Hao, L. Wang, X. Lu, X. Cao, S. Jia, Y. Hu, and Y. Xiao, Topological Atomic Spin Wave Lattices by Dissipative Couplings. \emph{Phys. Rev. Lett.} \textbf{130}, 153602 (2023).
%
\bibitem{Ion}
L. Ding, K. Shi, Q. Zhang, D. Shen, X. Zhang, and W. Zhang, Experimental Determination of $\mathcal{PT}$-Symmetric Exceptional Points in a Single Trapped Ion. \emph{Phys. Rev. Lett.} \textbf{126}, 083604 (2021).
%
\bibitem{Bardyn}
C.-E. Bardyn, \emph{et al.} Topology by dissipation. \emph{N. J. Phys.} \textbf{15}, 085001 (2013).
%
\bibitem{Leefmans}
C. Leefmans, \emph{et al.} Topological dissipation in a time-multiplexed photonic resonator network. \emph{Nat. Phys.} \textbf{18}, 442-449 (2022).
%
\bibitem{Yoshida}
T. Yoshida, and Y. Hatsugai, Bulk edge correspondence of classical difusion phenomena. \emph{Sci. Rep.} \textbf{11}, 888 (2021).
%
\bibitem{Parto}
M. Parto, C. Leefmans, J. Williams, F. Nori, and A. Marandi, Non-Abelian effects in dissipative photonic topological lattices. \emph{Nat. Commun.} \textbf{14}, 1440 (2023).
%
\end{thebibliography}
\end{document}